\documentstyle{article} 

\begin{document}       
\title{Momentum Distribution of the Hubbard Model}
\author{Girish S. Setlur 
 \\ The Institute of Mathematical Sciences}
\maketitle

\begin{abstract}
 Using the recently amended sea-boson method, we compute the 
 momentum distribution of the one-band Hubbard model
 in one and two spatial dimensions.
 We compute the asymptotic features of the momentum distribution explicitly 
 away from half filling for weak coupling in one
 and two dimensions. While the results are not exact by any means,
 they provide the exact asymptotics,
 namely they are able to reproduce the exponents
 obtained by Shulz in one dimension 
 obtained using Bethe ansatz. The corresponding results in
 more than one dimension are therefore as believeable. 
\end{abstract}

\section{Introduction}

 In the present article, we use the recently perfected sea-boson method
 \cite{Setlur1} \cite{Setlur2} \cite{Setlur3} to compute the
 momentum distribution of the one-band Hubbard model.
 The Hubbard model in two dimensions is thought to be important in
 understanding high-$ T_{c} $ superconductivity. Thus a 
 proper solution of this using the sea-boson method can shed new
 light on the question of breakdown of Fermi liquid behaviour in two
 dimensions from which one may decide whether or not the Hubbard model
 is adequate in describing high-$ T_{c} $ materials. 
 Furthermore, lattice Fermi gases are easy to solve using the sea-boson
 approach since the hamiltonian remains separable even after 
 invoking the repulsion attraction duality\cite{Setlur2}\cite{Setlur3}. 
 The anomalous exponent is computed
 for the 1d case and a comparison is made with the exact Bethe
 ansatz results of Lieb and Wu\cite{Lieb} as elaborated 
 by Shulz\cite{Shulz}. Unfortnately as in the usual bosonization
 in 1d\cite{Sumathi} we are unable to probe the strongly coupled
 ( large U ) Hubbard model since the
 velocities of the spinons becomes imaginary and the formalism breaks down. 
 However we take the plausible point of view, supported by earlier
 works\cite{Setlur3} , that says that breakdown of Fermi liquid theory
 (in any dimension) 
  comes from the infrared divergence of certain integrals in the sea-boson
 formalism and not from the strength of the coupling.  Thus it is sufficient
 to investigate the weakly coupled regime to ascertain whether or not
 Fermi liquid theory has broken down.  This point of view is further
 strengthened by the remarks of Shulz\cite{Shulz}.

 The Hubbard model is one of the most extensively studied models in 
 Condensed Matter Physics. The collection of reprints
 in the volume by Korepin and Essler\cite{Korepin} is particularly useful.
 Also the monumental works by Weng et.al.\cite{Weng} 
 which is a path integral approach to the 1d Hubbard model cannot be ignored.
 In more than one dimension, there are many attempts notable among them are
 by Metzner\cite{Metzner} and Fukuyama \cite{Fukuyama}. 

\section{Hubbard Model in One Dimension}

 Here we write down the one-band Hubbard model in one dimension
 in momentum space. To do this we must be careful since there are
 issues of backward scattering and umklapp process that play an important 
 role the latter especially at half-filling\cite{Emery}.
 Consider the model written in the usual manner. We assume that 
 we are in the thermodynamic limit so that we may assume that 
 the number of sites $ N_{a} \gg 1 $ is even. Also we set the hopping matrix
 element $ t = 1 $ and we work with units such that the lattice
 spacing $ a = 1 $. This means $ L = N_{a} $. 
\begin{equation}
H = -\sum^{m=N_{a}/2-1}_{m=-N_{a}/2 }
 \sum_{ \sigma } \mbox{       }(c^{\dagger}_{m \sigma }c_{m+1 \sigma}
 + c^{\dagger}_{m+1 \sigma }c_{m \sigma})
+  U \sum_{ m = -N_{a}/2 }^{ N_{a}/2-1}
 n_{m \uparrow} n_{m \downarrow}
\end{equation}
Here $  n_{ m \sigma } = c^{\dagger}_{ m \sigma }c_{ m \sigma } $.
\begin{equation}
c_{m \sigma} = \frac{1}{ \sqrt{ N_{a} }  }
\sum_{k} e^{ikm} \mbox{          }c_{k \sigma}
\end{equation}
\begin{equation}
c_{k \sigma} = \frac{1}{ \sqrt{ N_{a} }  }
\sum_{m = -N_{a}/2}^{N_{a}/2-1} e^{-ikm} \mbox{          }c_{m \sigma}
\end{equation}
Since we want periodic boundary conditions we must have,
\begin{equation}
c_{m+N_{a} \sigma} = c_{m \sigma }
\end{equation}
 This means that $ k \cdot N_{a} = 2 \pi \mbox{        }\nu $
 where $ \nu = 0, \pm 1, \pm 2, .... $. 
 We can also see that $ c_{k+2\pi, \sigma} = c_{k \sigma} $.
 This means that we may now assume that in the thermodynamic limit,
 the k's form a continuum and take on
 values in the interval $ k \in (-\pi,\pi) \equiv I$. 
 We may do so without any loss of generality.
 This perdiodicity leads to some novel processes such as umklapp  
 processes. To see this more clearly we write down the repulsion term 
 in momentum space.
 The hopping term is straightforward. To see this we write,
\begin{equation}
H_{kin} = -\frac{1}{N_{a}} \sum_{ \sigma }
\sum_{k,k^{'}}
 e^{ik} c^{\dagger}_{k^{'}\sigma} c_{k \sigma} f(k-k^{'}) 
 + H.c.
\end{equation} 
where,
\begin{equation}
f(k-k^{'}) = -2i\frac{ sin( (k-k^{'})N_{a}/2 ) }
{ 1 - e^{i(k-k^{'})} } = 0
\end{equation} 
if $ k-k^{'} \neq 0, \pm 2\pi, \pm 4\pi ,... $
and 
\begin{equation}
f(k-k^{'}) = N_{a}
\end{equation} 
if $ k-k^{'} = 0, \pm 2\pi, \pm 4\pi ,... $.
Since $ k,k^{'} \in (-\pi,\pi) $ there is no chance for
 $ k-k^{'} = \pm 2\pi $. Thus $ k - k^{'} = 0 $ is the only possibility.
 However in the repulsion term we find that such umklapp processes
 are indeed possible. The hopping term is therefore given by,
\begin{equation}
H_{kin} = \sum_{k \in I,\sigma} \epsilon_{k} c^{\dagger}_{k\sigma} c_{k\sigma}
\end{equation} 
and $  \epsilon_{k} = -2 \mbox{       }cos(k) $.
The interaction term is given by,
\begin{equation}
H_{U} = U \sum_{ m = -N_{a}/2 }^{ m = N_{a}/2 -1 }
n_{m \uparrow }n_{ m \downarrow }
\end{equation} 
In momentum space it is,
\begin{equation}
H_{U} = \frac{U}{N^{2}_{a}}
\sum_{k,k^{'}}\sum_{p,p^{'}}
c^{\dagger}_{ k^{'} \uparrow }
c_{ k \uparrow }c^{\dagger}_{p^{'} \downarrow } c_{p \downarrow }
f(k-k^{'}+p-p^{'})
\end{equation} 
\begin{equation}
f(k-k^{'}+p-p^{'}) = \sum_{m = -N_{a}/2}^{N_{a}/2 -1 }
e^{i m(k-k^{'}+p-p^{'})}
\end{equation} 
We can see that unless $ p = 0, \pm 2\pi, \pm 4\pi, .... $ 
 we must have $ f(p) \equiv 0 $.
Thus we may write,
\begin{equation}
f(k-k^{'}+p-p^{'}) = N_{a} ( \delta_{ k-k^{'}+p-p^{'}, 0 }
 +  \delta_{ k-k^{'}+p-p^{'}, 2\pi }
 + \delta_{ k-k^{'}+p-p^{'}, -2\pi } )
\end{equation} 
Since $ k,k^{'}, p, p^{'} \in (-\pi,\pi) $, there is no chance
 for $ k-k^{'}+p-p^{'} = \pm 4 \pi $ or anything higher
( since $ -4\pi < k-k^{'}+p-p^{'} < 4 \pi $ ). 
 However, $ k-k^{'}+p-p^{'} = \pm 2 \pi $ is perfectly possible, and this
 is precisely the umklapp process. Therefore we may write,
\begin{equation}
H_{U} = \frac{U}{N_{a}}
\sum_{k,k^{'}}\sum_{p,p^{'}}
c^{\dagger}_{ k^{'} \uparrow }
c_{ k \uparrow }c^{\dagger}_{p^{'} \downarrow } c_{p \downarrow }
( \delta_{k-k^{'} + p - p^{'},0}
 + \delta_{k-k^{'} + p - p^{'},2\pi}
 + \delta_{k-k^{'} + p - p^{'}, -2\pi} )
\end{equation} 
In other words,
\[
H_{U} = \frac{U}{N_{a}}
\sum_{k,p,q} c^{\dagger}_{ k + q \uparrow }
c_{ k \uparrow }c^{\dagger}_{p - q \downarrow } c_{p \downarrow }
+ \frac{U}{N_{a}}
\sum_{k,p,q}
c^{\dagger}_{ k + q \uparrow }
c_{ k \uparrow }c^{\dagger}_{p - q - G\downarrow } c_{p \downarrow }
\]
\begin{equation}
+ \frac{U}{N_{a}}
\sum_{k,p,q}
c^{\dagger}_{ k + q \uparrow }
c_{ k \uparrow }c^{\dagger}_{p - q + G\downarrow } c_{p \downarrow }
\end{equation} 
 The last two terms correspond to umklapp processes. Here $ G = 2 \pi $
 is the reciprocal lattice vector.
 Define the sea-displacement operator,
\begin{equation}
A_{ p\sigma }^{ p^{'}\sigma^{'} } =
 \frac{ n_{F}(p)(1 - n_{F}(p^{'})) }
{ \sqrt{ n_{ p\sigma } } }
c^{\dagger}_{ p\sigma }c_{  p^{'}\sigma^{'} }
\label{DEFN}
\end{equation}
 where $ n_{ p\sigma } = c^{\dagger}_{ p\sigma } \mbox{    }c_{ p\sigma } $
  and $ n_{F}(p) = \theta(\epsilon_{F}  - \epsilon_{p}) $.
 Eq.(~\ref{DEFN}) may be formally inverted and a formula for the number
 conserving product of two Fermi fields may be written down.
 In the RPA-sense we may regard the object
 $ A_{ p\sigma }^{ p^{'}\sigma^{'} } $ as being small in the sense that 
 the matrix respresentation of this operator in a suitably restricted 
 Hilbert space is sparse. Thus we can be content at including only 
 the leading terms. The formula for the off-diagonal product
 $ c^{\dagger}_{p\sigma}c_{p^{'}\sigma^{'}} $ for $ p \neq p^{'} $
 may be written down as follows.
\begin{equation}
 c^{\dagger}_{p\sigma}c_{p^{'}\sigma^{'}} \approx 
A_{ p\sigma }^{ p^{'}\sigma^{'} }
 + A_{ p^{'}\sigma^{'} }^{ \dagger p\sigma }
\end{equation}
 If $ p = p^{'} $ we have instead
 \footnote{ especially if $ n_{F}(p\uparrow) =  n_{F}(p\downarrow) $
 in other words if the number of up-spins is equal to the number
 of down spins $ M = M^{'} $ },
\begin{equation}
 c^{\dagger}_{p\sigma}c_{p\sigma^{'}} = 
\delta_{ \sigma, \sigma^{'} }
n_{F}(p) + \sum_{q \sigma_{1} }
A^{\dagger p \sigma}_{p-q \sigma_{1}}
A^{p \sigma^{'} }_{p-q \sigma_{1}}
 -  \sum_{q \sigma_{1} }
A^{\dagger p+q \sigma_{1} }_{p \sigma^{'}}
A^{p+q \sigma_{1} }_{p \sigma}
\end{equation}
 The interaction term with $ q = 0 $ may be omitted since this leads to
 a c-number. The terms with $ k = k^{'} \pm q $ are also omitted since
 they result in a hamiltonian that is quartic in the sea-bosons(see 
 later for some problems related to this). 
 Since these objects are small in the restricted Hilbert space we may
 ignore these terms.  The full hamiltonian may be written as the sum of
 two terms. The first corresponds to holons, the other corresponds to spinons.
 This is the first hint of the phenomenon of spin charge separation.  
 The full hamiltonian is $ H = H_{c} + H_{s} $ where,
\[
H_{c} = \sum_{k q} 
\epsilon_{k} \mbox{       }
A^{\dagger k\uparrow}_{k-q\uparrow}
A^{k\uparrow}_{k-q\uparrow}
 - \sum_{k q} 
\epsilon_{k} \mbox{       }
A^{\dagger k+q\uparrow}_{k\uparrow}
A^{k+q\uparrow}_{k\uparrow}
\]
\[
+ \sum_{k q} 
\epsilon_{k} \mbox{       }
A^{\dagger k\downarrow}_{k-q\downarrow}
A^{k\downarrow}_{k-q\downarrow}
 - \sum_{k q} 
\epsilon_{k} \mbox{       }
A^{\dagger k+q\downarrow}_{k\downarrow}
A^{k+q\downarrow}_{k\downarrow}
\]
\[
+ \frac{U}{N_{a}} \sum_{k,k^{'},q}
(A_{k+q \uparrow}^{k \uparrow} + 
A^{\dagger k+q \uparrow }_{k \uparrow})
(A_{k^{'}-q \downarrow}^{k^{'} \downarrow} + 
A^{\dagger k^{'}-q \downarrow }_{k^{'} \downarrow})
\]
\[
+ \frac{U}{N_{a}} \sum_{k,k^{'},q}
(A_{k+q \uparrow}^{k \uparrow} + 
A^{\dagger k+q \uparrow }_{k \uparrow})
(A_{k^{'}-q - G\downarrow}^{k^{'} \downarrow} + 
A^{\dagger k^{'}-q - G\downarrow }_{k^{'} \downarrow})
\]
\begin{equation}
+ \frac{U}{N_{a}} \sum_{k,k^{'},q}
(A_{k+q \uparrow}^{k \uparrow} + 
A^{\dagger k+q \uparrow }_{k \uparrow})
(A_{k^{'}-q + G\downarrow}^{k^{'} \downarrow} + 
A^{\dagger k^{'} - q + G\downarrow }_{k^{'} \downarrow})
\end{equation}
\[
H_{s} = \sum_{k q} 
\epsilon_{k} \mbox{       }
A^{\dagger k\uparrow}_{k-q\downarrow}
A^{k\uparrow}_{k-q\downarrow}
 - \sum_{k q} 
\epsilon_{k} \mbox{       }
A^{\dagger k+q\uparrow}_{k\downarrow}
A^{k+q\uparrow}_{k\downarrow}
\]
\[
+  \sum_{k q} 
\epsilon_{k} \mbox{       }
A^{\dagger k\downarrow}_{k-q\uparrow}
A^{k\downarrow}_{k-q\uparrow}
 - \sum_{k q} 
\epsilon_{k} \mbox{       }
A^{\dagger k+q\downarrow}_{k\uparrow}
A^{k+q\downarrow}_{k\uparrow}
\]
\[
- \frac{U}{N_{a}} \sum_{k,k^{'},q}
(A_{k+q \uparrow}^{k^{'} \downarrow} + 
A^{\dagger k+q \uparrow }_{ k^{'} \downarrow })
(A_{k^{'}-q \downarrow}^{k \uparrow} + 
A^{\dagger k^{'}-q \downarrow }_{ k \uparrow })
\]
\[
 - \frac{U}{N_{a}} \sum_{k,k^{'},q}
(A_{k+q \uparrow}^{k^{'} \downarrow} + 
A^{\dagger k+q \uparrow }_{ k^{'} \downarrow })
(A_{k^{'}-q - G\downarrow}^{k \uparrow} + 
A^{\dagger k^{'}-q -G \downarrow }_{ k \uparrow })
\]
\begin{equation}
  - \frac{U}{N_{a}} \sum_{k,k^{'},q}
(A_{k+q \uparrow}^{k^{'} \downarrow} + 
A^{\dagger k+q \uparrow }_{ k^{'} \downarrow })
(A_{k^{'}-q + G\downarrow}^{k \uparrow} + 
A^{\dagger k^{'}-q + G \downarrow }_{ k \uparrow })
\end{equation}
 $ A^{k^{'}\sigma}_{k\sigma} $ destroys an electron with
 some momentum and creates another with a different momentum without changing
 the spin. This means that this corresponds to charge fluctuations.
 $ A^{k^{'} {\bar{\sigma}} }_{k\sigma} $ destroys an electron with
 some momentum and creates another with a different momentum, and in addition
 changes the spin projection by $ \pm 1 $.
 This means that this corresponds to spin fluctuations.
 If we want the luxury of intepreting the objects
 $ A_{k\sigma}^{k^{'}\sigma^{'}} $ as being exact bosons obeying the
 canonical commutation rules :
\begin{equation}
[A_{k\sigma}^{k^{'}\sigma^{'}}, A_{k\sigma}^{\dagger k^{'}\sigma^{'}}]
 = n_{F}(k)(1 - n_{F}(k^{'}))
 \mbox{         }\theta(\pi-|k|) \theta(\pi-|k^{'}|) 
\end{equation}
 (and all other commutation rules involving any two of these objects are
 zero) then we must pay a price in that the momentum distribution is
 somewhat complicated, namely it is appropriately resummed in order to
 tame the infrared divergences that are known to be ubiquitous in
 one dimension.
 Thus the momentum distribution is given by,
\begin{equation}
<n_{k\sigma}> = n_{F}(k) 
F_{1}(k\sigma)
 + (1-n_{F}(k))
F_{2}(k\sigma)
\end{equation}
\begin{equation}
F_{1}(k\sigma) =
\frac{1}{2}\left( 1 + e^{ -2S^{0}_{B}(k\sigma) } \right)
\end{equation}
\begin{equation}
F_{2}(k\sigma) =
\frac{1}{2}\left( 1 - e^{ -2S^{0}_{A}(k\sigma) }  \right)
\end{equation}
\begin{equation}
S^{0}_{B}(k\sigma) =  \sum_{q_{1}\sigma_{1}}
 <G_{0}| A_{k\sigma}^{\dagger k+q_{1}\sigma_{1}} 
 A_{k\sigma}^{k+q_{1}\sigma_{1}} |G_{0}> 
\end{equation}
\begin{equation}
S^{0}_{A}(k\sigma)
 = \sum_{q_{1}\sigma_{1}}<G_{0}| A_{k-q_{1}\sigma_{1}}^{\dagger k\sigma} 
 A_{k-q_{1}\sigma_{1}}^{k\sigma} |G_{0}>
\end{equation}
 Here $ |G_{0}> $ is the ground state of $ H = H_{c} + H_{s} $. 
 In what follows, we assume that $ M = M^{'} $, that is, the total number
 of upspins is equal to the total number of downspins.

\section{Solution Away From Half-Filling}

We focus on the situation away from half-filling where the 
diagonalisation is simplest since we may ignore umklapp processes.
The holon part of the hamiltonian is given by,
\[
H_{c} = \sum_{k q} 
\epsilon_{k} \mbox{       }
A^{\dagger k\uparrow}_{k-q\uparrow}
A^{k\uparrow}_{k-q\uparrow}
 - \sum_{k q} 
\epsilon_{k} \mbox{       }
A^{\dagger k+q\uparrow}_{k\uparrow}
A^{k+q\uparrow}_{k\uparrow}
\]
\[
+ \sum_{k q} 
\epsilon_{k} \mbox{       }
A^{\dagger k\downarrow}_{k-q\downarrow}
A^{k\downarrow}_{k-q\downarrow}
 - \sum_{k q} 
\epsilon_{k} \mbox{       }
A^{\dagger k+q\downarrow}_{k\downarrow}
A^{k+q\downarrow}_{k\downarrow}
\]
\begin{equation}
+ \frac{U}{N_{a}} \sum_{k,k^{'},q}
(A_{k+q \uparrow}^{k \uparrow} + 
A^{\dagger k+q \uparrow }_{k \uparrow})
(A_{k^{'}-q \downarrow}^{k^{'} \downarrow} + 
A^{\dagger k^{'}-q \downarrow }_{k^{'} \downarrow})
\end{equation}
and the spinon part is given by,
\[
H_{s} = \sum_{k q} 
\epsilon_{k} \mbox{       }
A^{\dagger k\uparrow}_{k-q\downarrow}
A^{k\uparrow}_{k-q\downarrow}
 - \sum_{k q} 
\epsilon_{k} \mbox{       }
A^{\dagger k+q\uparrow}_{k\downarrow}
A^{k+q\uparrow}_{k\downarrow}
\]
\[
+  \sum_{k q} 
\epsilon_{k} \mbox{       }
A^{\dagger k\downarrow}_{k-q\uparrow}
A^{k\downarrow}_{k-q\uparrow}
 - \sum_{k q} 
\epsilon_{k} \mbox{       }
A^{\dagger k+q\downarrow}_{k\uparrow}
A^{k+q\downarrow}_{k\uparrow}
\]
\begin{equation}
- \frac{U}{N_{a}} \sum_{k,k^{'},q}
(A_{k+q \uparrow}^{k^{'} \downarrow} + 
A^{\dagger k+q \uparrow }_{ k^{'} \downarrow })
(A_{k^{'}-q \downarrow}^{k \uparrow} + 
A^{\dagger k^{'}-q \downarrow }_{ k \uparrow })
\end{equation}
 We use the equation of motion method to solve for the boson propagators.
 Since the details are tedious and the method has already been highlighted
 in our earlier work\cite{Setlur1}, we shall merely write down the final
 answers for the boson occupation.
\begin{equation}
\left< A^{\dagger k \sigma}_{k-q \sigma}(t^{+})
  A^{k \sigma}_{k-q \sigma}(t) \right>
 = \frac{ U^2 }{ N^2_{a} } \frac{i}{-i \beta}
\sum_{n} \frac{ P(q, iz_{n}) }{ \epsilon_{c}(q, iz_{n}) }
\frac{ [A_{k-q \sigma }^{k\sigma},
 A_{k-q \sigma }^{\dagger k\sigma}] }
{ (-iz_{n} - \epsilon_{k} + \epsilon_{k-q})^2 }
\end{equation}
\begin{equation}
 \left< A_{ k-q {\bar{ \sigma }} }^{\dagger k \sigma }(t^{+}) 
A_{k-q {\bar{ \sigma }} }^{ k \sigma }(t) \right> =
-\frac{ U }{ N_{a} } 
\frac{i}{-i \beta}
\sum_{n} \mbox{        }
\frac{1}{ \epsilon_{s}(q,iz_{n}) }
\frac{ [A^{k \sigma}_{k-q {\bar{\sigma}} },
 A^{\dagger k \sigma}_{k-q {\bar{\sigma}} }] }
{ (-iz_{n} - \epsilon_{ k } + \epsilon_{ k-q } )^2  } 
\end{equation}
\begin{equation}
P(q,iz_{n}) = 
\sum_{k}
 \frac{ [A_{k-q \sigma}^{k\sigma},A_{k-q \sigma}^{\dagger k\sigma}]  }
{ \left(-iz_{n} - \epsilon_{k} + \epsilon_{k-q} \right) }
+ \sum_{k}
 \frac{ [A_{k \sigma}^{k-q\sigma},A_{k\sigma}^{\dagger k-q\sigma}]  }
{ \left(iz_{n}- \epsilon_{k-q} +  \epsilon_{k} \right) }
\end{equation}
\begin{equation}
\epsilon_{c}(q, iz_{n}) =
1 - \frac{ U^2 }{ N^2_{a} }P^2(q,iz_{n})
\end{equation}
\begin{equation}
\epsilon_{s}(q,iz_{n}) = 1 + \frac{ U }{ N_{a} }P(q,iz_{n})
\end{equation}
 At absolute zero, the discrete sums become integrals and
 we have to evaluate the integral by transforming it into
 a contour integral in the complex plane such that  
 only zeros of the dielectric
 function ( and the pole of $ P $ ) are included.
 This means that we have to close the contour such that
 the zeros of the dielectric function are found at $ iz_{n} > 0 $. 
 Thus we may simplify the expressions as follows.
\[
\left< A^{\dagger k \sigma}_{k-q \sigma}(t^{+})
  A^{k \sigma}_{k-q \sigma}(t) \right>
 = \frac{ U^2 }{ N^2_{a} } 
 \frac{ 1 }{  \epsilon_{c}(q, \omega_{*}) 
 \frac{ d }{  d\omega } \vline_{\omega=\omega_{*} }
 P^{-1}(q,\omega)  }
\frac{ [A_{k-q \sigma }^{k\sigma},
 A_{k-q \sigma }^{\dagger k\sigma}] }
{ (-\omega_{*}(q) - \epsilon_{k} + \epsilon_{k-q})^2 }
\]
\begin{equation}
 + \frac{ U^2 }{ N^2_{a} } 
 \frac{ 1 }{
 P^{-1}(q, \omega_{c} )  \frac{ d }{  d\omega } 
\vline_{\omega = \omega_{c} } 
 \epsilon_{c}(q,\omega) }
\frac{ [A_{k-q \sigma }^{k\sigma},
 A_{k-q \sigma }^{\dagger k\sigma}] }
{ (-\omega_{c}(q) - \epsilon_{k} + \epsilon_{k-q})^2 }
\end{equation}
 Here $ \omega_{*} > 0 $ is the zero of $ P^{-1} $ and
 $ \omega_{c} > 0 $ is the zero of $ \epsilon_{c} $.
\begin{equation}
P^{-1}(q,\omega_{*}) = 0
\end{equation}
\begin{equation}
 \epsilon_{c}(q,\omega_{c}) = 0
\end{equation}
Similarly we may evaluate the other correlation function as,
\begin{equation}
 \left< A_{ k-q {\bar{ \sigma }} }^{\dagger k \sigma }(t^{+}) 
A_{k-q {\bar{ \sigma }} }^{ k \sigma }(t) \right> =
-\frac{ U }{ N_{a} } 
\frac{1}{ \frac{d}{ d \omega } \vline_{ \omega = \omega_{s} }
 \epsilon_{s}(q,\omega) }
\frac{ [A^{k \sigma}_{k-q {\bar{\sigma}} },
 A^{\dagger k \sigma}_{k-q {\bar{\sigma}} }] }
{ (-\omega_{s}(q) - \epsilon_{ k } + \epsilon_{ k-q } )^2  } 
\end{equation}
and $ \omega_{s} > 0  $ is the zero of $ \epsilon_{s} $.
\begin{equation}
 \epsilon_{s}(q,\omega_{s}) = 0
\end{equation}
 Since we are interested only in the asymptotics we focus on the
 small $ |q| $ regime. In this regime,
\begin{equation}
P(q,\omega) \approx \frac{L}{2 \pi}
 \frac{ 4 \mbox{    }sin(k_{F}) \mbox{        }q^2 }
{ \left( \omega^2 - 4 sin^2(k_{F}) q^2 \right) } 
\end{equation}
 Since $ \epsilon_{c}(q, \omega_{*}) = \infty $ we may ignore
 this contribution. Let us now enumerate all the zeros
 of the dielectric function.
\begin{equation}
\epsilon_{c}(q,\omega) \equiv 1 -  \frac{ U^2 }{ (2 \pi)^2 }
 \frac{ (4 \mbox{    }sin(k_{F}) \mbox{        }q^2)^2 }
{ \left( \omega_{c}^2 - 4 sin^2(k_{F}) q^2 \right)^2 } = 0
\end{equation}
\begin{equation}
\epsilon_{s}(q,\omega) \equiv 1 + \frac{ U }{ (2 \pi) }
 \frac{ (4 \mbox{    }sin(k_{F}) \mbox{        }q^2) }
{ \left( \omega_{s}^2 - 4 sin^2(k_{F}) q^2 \right) } = 0
\end{equation}
 There is only one zero of $ \epsilon_{s} $ namely, 
\begin{equation}
\omega_{s}(q) = v_{F} |q| 
\left( 1 - \frac{ U }{ \pi v_{F} } \right)^{\frac{1}{2}}
 = v_{s} |q|
\end{equation}
where $ v_{F} = 2 \mbox{      }sin(k_{F}) $. Thus we may write
 the spinon hamiltonian in terms of the elementary
 excitations of the interacting system as,
\begin{equation}
H_{s} = \sum_{q} \omega_{s}(q) d^{\dagger}_{s}(q)
d_{s}(q) 
\end{equation}
and for the holons,
\begin{equation}
\omega_{c,1}(q) = v_{F} |q| 
\left( 1 - \frac{ U }{\pi v_{F} } \right)^{\frac{1}{2}}
 = v_{c,1} |q|
\end{equation}
\begin{equation}
\omega_{c,2}(q) = v_{F} |q| 
\left( 1 + \frac{ U }{ \pi  v_{F} } \right)^{\frac{1}{2}}
 = v_{c,2} |q|
\end{equation}
\begin{equation}
H_{c} = \sum_{q} \omega_{c,1}(q) d^{\dagger}_{c,1}(q)
d_{c,1}(q) +  \sum_{q} \omega_{c,2}(q)
 d^{\dagger}_{c,2}(q)d_{c,2}(q)
\end{equation}
 It appears that these formulas are valid only for weak coupling.
 For strong coupling we shall have to find some other way. For now
 we focus only on weak coupling. 
\[
\left< A^{\dagger k \sigma}_{k-q \sigma}(t^{+})
  A^{k \sigma}_{k-q \sigma}(t) \right>
 = \frac{ U^2 }{ N^2_{a} } 
 \sum_{j=1,2} \frac{ 1 }{
 P^{-1}(q, \omega_{c,j} )  \frac{ d }{  d\omega } 
\vline_{\omega = \omega_{c,j} } 
 \epsilon_{c}(q,\omega) }
\frac{ [A_{k-q \sigma }^{k\sigma},
 A_{k-q \sigma }^{\dagger k\sigma}] }
{ (-\omega_{c,j}(q) - \epsilon_{k} + \epsilon_{k-q})^2 }
\]
\begin{equation}
 = \sum_{j=1,2}  \frac{ U^2 }{ 4 \pi N_{a} } |q|
  \frac{ v_{F} }{ v_{c,j} }
\frac{ [A_{k-q \sigma }^{k\sigma},
 A_{k-q \sigma }^{\dagger k\sigma}] }
{ (-v_{c,j}|q| - v_{F} \mbox{     }q  \mbox{       }sgn(k) )^2 }
\end{equation}
\begin{equation}
 \left< A_{ k-q {\bar{ \sigma }} }^{\dagger k \sigma }(t^{+}) 
A_{k-q {\bar{ \sigma }} }^{ k \sigma }(t) \right> =
-\frac{ U }{ N_{a} } 
\frac{ |q| }{ -(2 \pi/U) (v_{s}/v_{F}) }
\frac{ [A^{k \sigma}_{k-q {\bar{\sigma}} },
 A^{\dagger k \sigma}_{k-q {\bar{\sigma}} }] }
{ (-v_{s}|q| - v_{F} \mbox{       }q \mbox{      }sgn(k) )^2  } 
\end{equation}
Therefore for $ k $ close to $ +k_{F} $ we may write
 ( $ k = k_{F} + x $ ),
\[
S^{0}_{A}(k\sigma) = \sum_{q} < A^{\dagger k \sigma }_{k-q \sigma}
A^{k\sigma}_{k-q \sigma} >
 +  \sum_{q} < A^{\dagger k \sigma }_{k-q {\bar{ \sigma }} }
A^{k\sigma}_{k-q {\bar{ \sigma }} } >
\]
\begin{equation}
 = \int_{ |x| }^{ \Lambda }
 \frac{ dq }{q} \mbox{       }\sum_{j=1,2}  \frac{ U^2 }{ 8 \pi^2 } 
  \frac{ v_{F} }{ v_{c,j} }
\frac{ 1 }{ (v_{c,j} + v_{F})^2 }
 +  \int_{ |x| }^{ \Lambda }
 \frac{ dq }{q} \mbox{        }
\frac{ U^2 }{ 4 \pi^2 }
\frac{ v_{F} }{ v_{s} }
\frac{ 1 }
{ (v_{s} + v_{F} )^2  }  = \frac{\gamma }{2}
 Log \left[ \frac{ \Lambda }{|x|} \right]
\end{equation}
The anomalous exponent is then given by,
\begin{equation}
\gamma = \sum_{j=1,2}  \frac{ U^2 }{ 4 \pi^2 } 
  \frac{ v_{F} }{ v_{c,j} }
\frac{ 1 }{ (v_{c,j} + v_{F})^2 }
 + \frac{ U^2 }{ 2\pi^2 }
\frac{ v_{F} }{ v_{s} }
\frac{ 1 }
{ (v_{s} + v_{F} )^2  }
\label{ANOM} 
\end{equation}
 The velocities of spinons and holons are consistent with the ones
 obtained by traditional bosonization methods. The anomalous exponent is 
 proportional to $ U^2 $ for small $ U $ both in the work
 by Shulz\cite{Shulz} and our work.
 The momentum distribution is then
 given by ( $ ||k| - k_{F}| < \Lambda $ ),
\begin{equation}
{\bar{n}}_{k} = \frac{1}{2} - \frac{1}{2} 
 sgn(|k|-k_{F}) \left( \frac{ ||k|-k_{F}| }{ \Lambda } \right)^{\gamma} 
\end{equation}
  For a detailed comparison with the results of Shulz\cite{Shulz} 
  we note that Shulz points out that the charge stiffness may be written
  as a power series in $ U $ for small $ U $ as
  $ K_{ \rho } = 1 - U/(\pi v_{F}) + ... $. This means that we may
 read off a formula for the anomalous exponent
 $ \gamma = (K_{\rho} + 1/K_{\rho} - 2)/4 \approx U^2/(2 \pi v_{F})^2 $. 
  From Eq.(~\ref{ANOM})
  we find that since $ v_{s} \approx v_{c,1} \approx v_{c,2} \approx v_{F} $, 
  we may conclude that $ \gamma \approx U^2/(4 \pi^2 v^2_{F}) $. 
  This agrees exactly with the Shulz result above.  In the large U-limit,
  the velocities of the spinons becomes imaginary and we are unable
  to make progress. This means that the cubic and quartic terms
  that the above approach ignores are now relevant. That is, their inclusion
  alters the value of the exponents. At half-filling again there seem
  to be some difficulties which we have been unable to overcome. Perhaps
  future publications will address these issues. 

\section{ Momentum Distribution in Two Dimensions }

 The main advantage of the sea-boson approach is the ease with which
 one may generalise the above results to more than one dimension by
 simply promoting the wavenumbers to wave-vectors.
\begin{equation}
\left< A^{\dagger {\bf{k}} \sigma}_{ {\bf{k}}-{\bf{q}} \sigma}(t^{+})
  A^{{\bf{k}} \sigma}_{ {\bf{k}}-{\bf{q}} \sigma}(t) \right>
 = \frac{ U^2 }{ N^2_{a} } \frac{i}{-i \beta}
\sum_{n} \frac{ P({\bf{q}},iz_{n}) }{ \epsilon_{c}({\bf{q}}, iz_{n}) }
\frac{ [A_{ {\bf{k}}-{\bf{q}} \sigma }^{ {\bf{k}} \sigma},
 A_{ {\bf{k}} - {\bf{q}} \sigma }^{\dagger {\bf{k}}\sigma}] }
{ (-iz_{n} - \epsilon_{ {\bf{k}} } + \epsilon_{ {\bf{k}}-{\bf{q}} })^2 }
\end{equation}
\begin{equation}
 \left< A_{ {\bf{k}}-{\bf{q}} {\bar{ \sigma }}
 }^{\dagger {\bf{k}} \sigma }(t^{+}) 
A_{{\bf{k}}-{\bf{q}} {\bar{ \sigma }} }^{ {\bf{k}} \sigma }(t) \right> =
-\frac{ U }{ N_{a} } 
\frac{i}{-i \beta}
\sum_{n} \mbox{        }
\frac{1}{ \epsilon_{s}( {\bf{q}}, iz_{n}) }
\frac{ [A^{{\bf{k}} \sigma}_{{\bf{k}}-{\bf{q}} {\bar{\sigma}} },
 A^{\dagger {\bf{k}} \sigma}_{{\bf{k}}-{\bf{q}} {\bar{\sigma}} }] }
{ (-iz_{n} - \epsilon_{ {\bf{k}} } + \epsilon_{ {\bf{k}}-{\bf{q}} } )^2  } 
\end{equation}
\begin{equation}
P({\bf{q}},iz_{n}) = 
\sum_{k}
 \frac{ [A_{{\bf{k}}-{\bf{q}} \sigma}^{k\sigma},A_{{\bf{k}}-{\bf{q}} 
\sigma}^{\dagger {\bf{k}} \sigma}]  }
{ \left(-iz_{n} - \epsilon_{ {\bf{k}} } 
+ \epsilon_{{\bf{k}}-{\bf{q}}} \right) }
+ \sum_{ {\bf{k}} }
 \frac{ [A_{{\bf{k}} \sigma}^{{\bf{k}}-{\bf{q}}\sigma},
A_{ {\bf{k}} \sigma}^{\dagger {\bf{k}}-{\bf{q}}\sigma}]  }
{ \left(iz_{n}- \epsilon_{{\bf{k}}-{\bf{q}}} +  \epsilon_{ {\bf{k}} } \right) }
\end{equation}
\begin{equation}
\epsilon_{c}({\bf{q}},iz_{n}) =
1 - \frac{ U^2 }{ N^2_{a} }P^2({\bf{q}},iz_{n})
\end{equation}
\begin{equation}
\epsilon_{s}({\bf{q}},iz_{n}) = 1 + \frac{ U }{ N_{a} }P({\bf{q}},iz_{n})
\end{equation}
 For evaluating the low energy, long-wavelength limit of the above expressions,
 we may ignore the non-zero lattice spacing and instead focus on the
 continuum limit (which is valid for $ k_{F} \ll 1 $ since we have
 set $ a = 1 $) where the expressions for the RPA-polarization have
 already been derived by Stern\cite{Stern}(here $ 2m = 1$).
\begin{equation}
Re[P]({\bf{q}},\omega) = -\frac{ m k_{F} A }{2 \pi |{\bf{q}}| }
\{ \frac{ | {\bf{q}}| }{ k_{F} } 
- C_{-} \left[  \left( \frac{ |{\bf{q}}| }{2k_{F}}
 -  \frac{ m \omega }{ k_{F} |{\bf{q}}| }   \right)^2 - 1 \right]^{\frac{1}{2}}
 -  C_{+} \left[  \left( \frac{ |{\bf{q}}| }{2k_{F}}
 +  \frac{ m \omega }{ k_{F} |{\bf{q}}| }  \right)^2 - 
1 \right]^{\frac{1}{2}} \}
\end{equation}
\begin{equation}
Im[P]({\bf{q}},\omega) = -\frac{ m k_{F} A }{ 2 \pi |{\bf{q}}| }
\{ D_{-} \left[ 1 - \left[ \frac{ |{\bf{q}}| }{2k_{F}} 
- \frac{ m \omega }{ k_{F} |{\bf{q}}| }
 \right]^2
 \right]^{\frac{1}{2}}
- D_{+} \left[ 1 - \left[ \frac{ |{\bf{q}}| }{2k_{F}} 
+ \frac{ m \omega }{ k_{F} |{\bf{q}}| } 
 \right]^2
 \right]^{\frac{1}{2}} \}
\end{equation}
where,
\begin{equation}
C_{ \pm }  =  sgn \left[ \frac{ |{\bf{q}}| }{2k_{F}} 
\pm \frac{ m \omega }{ k_{F} |{\bf{q}}| } \right],
\mbox{         }
D_{ \pm }  = 0, \mbox{       }
 \left|  \frac{ |{\bf{q}}| }{2k_{F}} \pm
 \frac{ m \omega }{ k_{F} |{\bf{q}}| } \right| > 1 
\end{equation}
\begin{equation}
C_{ \pm }  = 0, \mbox{       } D_{ \pm } = 1, \mbox{       } 
 \left|  \frac{ |{\bf{q}}| }{2k_{F}} \pm 
\frac{ m \omega }{ k_{F} |{\bf{q}}| } \right| < 1 
\end{equation}
\begin{equation}
Re[P]({\bf{q}},\omega) \approx -\frac{ m A }{ 2 \pi }
\{ 1 - \frac{ \omega  }
{ \sqrt{ \omega^2 - v_{F}^2 {\bf{q}}^2 } } \}
\end{equation}
\begin{equation}
0 = 1 + \frac{ m U }{ 2 \pi }
\{ 1 - \frac{ \omega  }
{ \sqrt{ \omega^2 - v_{F}^2 {\bf{q}}^2 } } \}
\end{equation}
for $ \omega > v_{F} | {\bf{q}} | $.
\begin{equation}
v_{eff} = v_{F} \mbox{         }
\frac{ \left( \frac{ 1 + \frac{ U }{ 4 \pi } }
{  \frac{ U }{ 4 \pi } } \right) }
{  \sqrt{ \left( \frac{ 1 + \frac{ U }{ 4 \pi } }
{  \frac{ U }{ 4 \pi } } \right)^2 - 1 } }
\end{equation}
\begin{equation}
\frac{ \partial }{ \partial \omega }
 \mbox{       }Re[P]({\bf{q}},\omega) \approx 
 - \frac{ A m v^2_{F} {\bf{q}}^2 }
{ 2 \pi (\omega^2 -  v^2_{F} {\bf{q}}^2 )^{3/2} }
\end{equation}
This means,
\begin{equation}
\left< A^{\dagger {\bf{k}} \sigma}_{ {\bf{k}}-{\bf{q}} \sigma}(t^{+})
  A^{ {\bf{k}} \sigma}_{ {\bf{k}} - {\bf{q}} \sigma}(t) \right>
 = \frac{ U^2 }{ N^2_{a} } 
 \frac{ 1 }{
 P^{-1}( {\bf{q}}, \omega_{c} )  \frac{ d }{  d\omega } 
\vline_{\omega = \omega_{c} } 
 \epsilon_{c}( {\bf{q}},\omega) }
\frac{ [A_{ {\bf{k}}-{\bf{q}} \sigma }^{ {\bf{k}} \sigma},
 A_{ {\bf{k}} - {\bf{q}} \sigma }^{\dagger {\bf{k}} \sigma}] }
{ (-\omega_{c}( {\bf{q}} ) - \epsilon_{ {\bf{k}} } 
+ \epsilon_{ {\bf{k}} - {\bf{q}} })^2 }
\end{equation}
\begin{equation}
\left< A^{\dagger {\bf{k}} \sigma}_{ {\bf{k}}-{\bf{q}} {\bar{\sigma}} }(t^{+})
  A^{ {\bf{k}} \sigma}_{ {\bf{k}} - {\bf{q}} {\bar{\sigma}} }(t) \right> = 0
\end{equation}
 and $ \omega_{c}({\bf{q}}) = v_{eff}|{\bf{q}}| $. For a triangular 
 lattice the energy dispersion may be written as, $ \epsilon_{ {\bf{k}} } =
 -2 \mbox{      }[ cos(k_{x}) + 2 \mbox{     } cos(k_{x}/2) \mbox{       }
cos( \sqrt{3} k_{y}/2) ] $.
 Let us assume that $ k_{F} a \ll 1 $ 
 (since we have set $ a = 1 $, we must have $ k_{F} \ll 1 $)
 in which case the Fermi surface is a circle,
\[
\left< A^{\dagger {\bf{k}} \sigma}_{ {\bf{k}}-{\bf{q}} \sigma}(t^{+})
  A^{ {\bf{k}} \sigma}_{ {\bf{k}} - {\bf{q}} \sigma}(t) \right>
 = -\frac{ 1 }{
 2 \frac{ d }{  d\omega } 
\vline_{\omega = \omega_{c} } Re[P]({\bf{q}},\omega) }
\frac{ [A_{ {\bf{k}}-{\bf{q}} \sigma }^{ {\bf{k}} \sigma},
 A_{ {\bf{k}} - {\bf{q}} \sigma }^{\dagger {\bf{k}} \sigma}] }
{ (-\omega_{c}( {\bf{q}} ) - \epsilon_{ {\bf{k}} } 
+ \epsilon_{ {\bf{k}} - {\bf{q}} })^2 }
\]
\[
= \frac{ \pi |{\bf{q}}| (v_{eff}^2 -  v^2_{F})^{3/2} }{ A m v^2_{F} }
\frac{ [A_{ {\bf{k}}-{\bf{q}} \sigma }^{ {\bf{k}} \sigma},
 A_{ {\bf{k}} - {\bf{q}} \sigma }^{\dagger {\bf{k}} \sigma}] }
{ (-v_{eff}|{\bf{q}}| - {\bf{k}}^2 
+ ({\bf{k}} - {\bf{q}})^2 )^2  }
\]
\[
\approx \frac{ \pi |{\bf{q}}| (v_{eff}^2 -  v^2_{F})^{3/2} }{ A m v^2_{F} }
\frac{ \theta(k_{F}-|{\bf{k}}-{\bf{q}}|) \mbox{     }
\theta(|{\bf{k}}|-k_{F}) }
{ (-v_{eff}|{\bf{q}}|  
- 2 {\bf{k}} \cdot {\bf{q}} + {\bf{q}}^2 )^2  }
\]
\begin{equation}
\approx \frac{ \pi |{\bf{q}}| (v_{eff}^2 -  v^2_{F})^{3/2} }{  A m v^2_{F} }
\frac{ \theta(k^2_{F}-{\bf{k}}^2 
+ 2 |{\bf{k}}| |{\bf{q}}| cos(\alpha) - {\bf{q}}^2 ) \mbox{     }
\theta(|{\bf{k}}|-k_{F}) }
{ (-v_{eff}|{\bf{q}}|  
- 2 |{\bf{k}}| |{\bf{q}}| cos(\alpha) + {\bf{q}}^2 )^2  }
\end{equation}
\[
\sum_{ {\bf{q}} }
 \left< A^{\dagger {\bf{k}} \sigma}_{ {\bf{k}}-{\bf{q}} \sigma}(t^{+})
  A^{ {\bf{k}} \sigma}_{ {\bf{k}} - {\bf{q}} \sigma}(t) \right>
 = 2 \int^{\pi/2}_{0}   d \alpha \mbox{        }
\int^{\infty}_{0} dq  \mbox{        }
\frac{ (v_{eff}^2 -  v^2_{F})^{3/2} }{ 2 \pi v^2_{F} }
\frac{ \theta(k^2_{F}-k^2 
+ 2 k_{F} q cos(\alpha) - q^2 ) \mbox{     }
\theta(|{\bf{k}}|-k_{F}) }
{ (-v_{eff} 
- 2 k_{F} cos(\alpha) + q )^2  }
\]
\begin{equation}
 = \frac{ (v_{eff}^2 -  v^2_{F})^{3/2} }{ \pi v^2_{F} }
\left( \frac{ \pi }{ 2 v_{eff} }
 - 2 \frac{ ArcTan \left[ 
\frac{ v_{eff}-v_{F} }{ \sqrt{ v_{eff}^2 - v_{F}^2 } } \right] }
{ \sqrt{ v_{eff}^2 - v_{F}^2 } } \right) 
\end{equation}
\begin{equation}
Z_{F} = exp \left[  -\frac{ (v_{eff}^2 -  v^2_{F})^{3/2} }{ \pi v^2_{F} }
\left( \frac{ \pi }{ v_{eff} }
 - \frac{ 4 \mbox{    }ArcTan \left[ 
\frac{ v_{eff}-v_{F} }{ \sqrt{ v_{eff}^2 - v_{F}^2 } } \right] }
{ \sqrt{ v_{eff}^2 - v_{F}^2 } } \right)  \right]
\end{equation}
For $ U \rightarrow +\infty $, we have,
\begin{equation}
v_{eff} \approx v_{F} \mbox{         } 
\left( \frac{ U }{ 8 \pi } \right)^{\frac{1}{2}} 
\end{equation}
In this limit,
\begin{equation}
\sum_{ {\bf{q}} }
 \left< A^{\dagger {\bf{k}} \sigma}_{ {\bf{k}}-{\bf{q}} \sigma}(t^{+})
  A^{ {\bf{k}} \sigma}_{ {\bf{k}} - {\bf{q}} \sigma}(t) \right>
 = \frac{ v_{eff} }{ 2 v_{F} }
\end{equation}
\begin{equation}
Z_{F}(U \rightarrow +\infty) \approx
 exp \left[ - \left( \frac{ U }{ 8 \pi } \right)^{\frac{1}{2}}  \right]
\end{equation}
For $ U \rightarrow +0 $, we have,
\begin{equation}
v_{eff} \approx v_{F} \mbox{         } 
\left( 1 + \frac{ U^2 }{ 32 \pi^2 } \right) 
\end{equation}
\begin{equation}
 Z_{F}(U \rightarrow 0^{+}) \approx
 1 - U^3 \mbox{       }\left( \frac{ \pi-2 }{ 64 \pi^4 } \right) 
\end{equation}
 The above results are valid when it is legitimate to approximate the shape
 of the Fermi surface by a circle. Strictly speaking therefore it is valid
 only for $ k_{F} \ll 1 $.
 Also the $ U \rightarrow +\infty $ result must be taken
 with a grain of salt since the corresponding 1d case has not worked out.
 Just as the anomalous exponent saturates to a value of $ 1/8 $
 according to Bethe ansatz and also hopefully according to the sea-boson
 method provided we include the anharmonic terms, we may suspect
 that the quasiparticle residue also saturates (to a value close to unity)
 when these anharmonic terms are included.
 This is completely consistent with the results of Castro-Neto and Fradkin
 (who predicted a value of $ 0.78 $ I think for the smallest value of the
 quasiparticle residue). However it is not clear to the author if their
 results were due to a formalism more powerful than the sea-boson method,
 or because of a clever use of intuition. In particular one has to see if
 the traditional bosonization method can predict the exponent of $ 1/8 $
 in 1d, without relying on the Bethe ansatz results.
 The small $ U $ results are probably reliable
 though one has to check this several times more.
 The quasiparticle residue seems to deviate from unity by a term proportional
 to $ U^3 $ rather than $ U^2 $.

\section{ Full Propagator }

 To compute say the density of states, we have to compute the dynamical
 one-particle Green function. For this we have to complete the
 construction of the field operator started in the Appendix. In particular
 we have to find a formula for $ R({\bf{p}} \sigma) $. 
\begin{equation}
c_{ {\bf{p}} \sigma } \equiv e^{ 
\sum_{ {\bf{q}} \sigma_{1} }
\left( -A_{ {\bf{p}} \sigma }
^{ \dagger {\bf{p}}+{\bf{q}}\sigma_{1} }
 + A_{ {\bf{p}} + {\bf{q}} \sigma_{1} }
^{ {\bf{p}} \sigma }  \right)
\mbox{      }T_{ {\bf{q}} }({\bf{p}}) }
\mbox{        }R({\bf{p}} \sigma)
\end{equation}
\begin{equation}
c^{\dagger}_{ {\bf{p}} \sigma } \equiv e^{ 
\sum_{ {\bf{q}} \sigma_{1} }
\left( -A_{ {\bf{p}} \sigma }
^{ {\bf{p}}+{\bf{q}}\sigma_{1} }
 + A^{\dagger {\bf{p}} \sigma }_{ {\bf{p}} + {\bf{q}} \sigma_{1} }
 \right)
\mbox{      }T_{ {\bf{q}} }({\bf{p}}) }
\mbox{        }R^{*}({\bf{p}} \sigma)
\end{equation}
 The main problem with these formulas is that it is not possible
 to use the Baker-Hausdorff theorem since the commutator of the two
 exponents does not commute with either even at equal times.
 Therefore it is not possible to systematically deduce $ R $ by making
 contact with the free theory. However we may suspect that perhaps,
\begin{equation}
R({\bf{p}} \sigma) = n_{F}({\bf{p}}) \mbox{        }
e^{i N^{0} \mbox{       }\xi({\bf{p}}\sigma) }
\end{equation}
as usual $ n_{F}({\bf{p}}) = \theta(k_{F}-|{\bf{p}}|) $.
Here $ \xi({\bf{p}}\sigma) $ is an arbitrary non-constant function
 (c-number)
of the arguments such that,
\begin{equation}
e^{-i N^{0} \mbox{       }\xi({\bf{p}}\sigma) }
e^{i N^{0} \mbox{       }\xi({\bf{p}}^{'}\sigma^{'}) }
 \approx \delta_{ {\bf{p}}, {\bf{p}}^{'} } \mbox{     }
\delta_{ \sigma, \sigma^{'} }
\end{equation}
Therefore,
\begin{equation}
\left< c^{\dagger}_{ {\bf{p}}^{'} \sigma^{'} }(t) 
c_{ {\bf{p}} \sigma }(0) \right>
= \left< e^{ 
\sum_{ {\bf{q}} \sigma_{1} }
\left( -A_{ {\bf{p}}^{'} \sigma^{'} }
^{ {\bf{p}}^{'}+{\bf{q}}\sigma_{1} }(t)
 + A^{\dagger {\bf{p}}^{'} \sigma^{'} }_{ {\bf{p}}^{'} + {\bf{q}}
 \sigma_{1} }(t)
 \right)
\mbox{      }T_{ {\bf{q}} }({\bf{p}}^{'}) }
 e^{ 
\sum_{ {\bf{q}} \sigma_{1} }
\left( -A_{ {\bf{p}} \sigma }
^{ \dagger {\bf{p}}+{\bf{q}}\sigma_{1} }
 + A_{ {\bf{p}} + {\bf{q}} \sigma_{1} }
^{ {\bf{p}} \sigma }  \right)
\mbox{      }T_{ {\bf{q}} }({\bf{p}}) }
\mbox{        }R^{*}({\bf{p}}^{'} \sigma^{'})
\mbox{        }R({\bf{p}} \sigma) \right>
\end{equation}
For reasons already mentioned this is not possible to simplify further.

\section{Conclusions}

 We have computed the asymptotic features of the momentum distribution of
 the one-band Hubbard model in one and two spatial dimensions. In the
 one dimensional case, we obtain a Luttinger liquid with an anomalous 
 exponent identical to the one
 obtained by Shulz\cite{Shulz} from Bethe ansatz. 
 This is valid for weak coupling only. We are unable to extend the results
 to strong coupling. In the two dimensional case, we obtain a Landau Fermi
 liquid with quasiparticle residue deviating from unity by a term proportional
 to the cube of $ U $. Using the remarks of Shulz, we may conclude
 that the two dimensional case is always a Landau Fermi liquid.

\section{ Appendix : The Field Operator }

 In the past, we have focussed on the density phase variable ansatz (DPVA) as
 the main way of ascertaining the connection between the
 field operator and the sea-displacements. This is done in a two step 
 process. The first step involves expressing
 the fields in terms of currents and
 densities. Next we relate the currents and
 densities to the sea-displacements. Unfortunately this program has faced
 difficulties since it appears now that the phase functional is extremely
 nontrivial thus making the approach practically useless. Here we instead
 focus on trying to pin down the field in terms of the sea-displacements
 directly. This approach is better since we now have a definition for the
 sea-displacements in terms of the Fermi fields.
 Consider the commutators(we have dropped the square roots of the
 number operators in the denominator),
\begin{equation}
[c_{ {\bf{p}} \sigma }, A_{ {\bf{k}}\sigma_{1} }({\bf{q}}\sigma_{2})]
 \approx \delta_{ {\bf{p}} + {\bf{q}}/2, {\bf{k}} }
n_{F}({\bf{p}})(1-n_{F}({\bf{p}}+{\bf{q}}))
\mbox{      }T_{ {\bf{q}} }({\bf{p}}) c_{ {\bf{p}} \sigma }
\mbox{         }\delta_{ \sigma, \sigma_{1} }
\label{COMM1}
\end{equation}
\begin{equation}
[c_{ {\bf{p}} \sigma }, A^{\dagger}_{ {\bf{k}} \sigma_{1} }
({\bf{q}}\sigma_{2})]
 \approx \delta_{ {\bf{p}}, {\bf{k}}+{\bf{q}}/2 }
n_{F}( {\bf{p}}-{\bf{q}} )( 1 - n_{F}({\bf{p}}) )
\mbox{      }T_{ -{\bf{q}} }({\bf{p}}) c_{ {\bf{p}} }
\delta_{ \sigma, \sigma_{2} }
\label{COMM2}
\end{equation}
 where $ T_{ {\bf{q}} }({\bf{p}}) \equiv exp({\bf{q}}.\nabla_{ {\bf{p}} }) $.
 We may now exponentiate the commutation rules in Eq.(~\ref{COMM1}) 
 and  Eq.(~\ref{COMM2}) and arrive at, 
\begin{equation}
c_{ {\bf{p}} \sigma } \equiv e^{ 
\sum_{ {\bf{q}} \sigma^{'} }
\left( -A^{\dagger}_{ {\bf{p}} + {\bf{q}}/2 \sigma }
({\bf{q}}\sigma^{'})
 + A_{ {\bf{p}} + {\bf{q}}/2 \sigma^{'} }
(-{\bf{q}}\sigma) \right)
\mbox{      }T_{ {\bf{q}} }({\bf{p}}) }
\mbox{        }R({\bf{p}} \sigma)
\label{CP}
\end{equation}
\begin{equation}
c^{\dagger}_{ {\bf{p}} \sigma } \equiv e^{ 
\sum_{ {\bf{q}} \sigma^{'} }
\left( -A_{ {\bf{p}} + {\bf{q}}/2 \sigma }
({\bf{q}}\sigma^{'})
 + A^{\dagger}_{ {\bf{p}} + {\bf{q}}/2 \sigma^{'} }
(-{\bf{q}}\sigma) \right)
\mbox{      }T_{ {\bf{q}} }({\bf{p}}) }
\mbox{        }R^{*}({\bf{p}} \sigma)
\label{CPDAG}
\end{equation}
 Here $ R({\bf{p}}\sigma) $ is a c-number
 function that commutes with everything.
 To verify the commutation rules we proceed as follows.
 The exponential is interpreted as a power series expansion.
 In the expansion, the translation operator translates all the 
 $ {\bf{p}}'s $ to its right by an amount $ {\bf{q}} $ and not just the
 $ {\bf{p}} $ in $ R({\bf{p}}\sigma) $.
\[
[c_{ {\bf{p}} \sigma }, A_{ {\bf{k}} \sigma_{1} }({\bf{q}}\sigma_{2})]
 = e^{ 
\sum_{ {\bf{q}} \sigma^{'} }
\left( -A^{\dagger}_{ {\bf{p}} + {\bf{q}}/2 \sigma }
({\bf{q}}\sigma^{'})
 + A_{ {\bf{p}} + {\bf{q}}/2 \sigma^{'} }
(-{\bf{q}}\sigma) \right)
\mbox{      }T_{ {\bf{q}} }({\bf{p}}) }  A_{ {\bf{k}} \sigma_{1} }({\bf{q}}\sigma_{2})
\mbox{        }R({\bf{p}} \sigma)
\]
\[
 - A_{ {\bf{k}} \sigma_{1} }({\bf{q}}\sigma_{2})
 e^{ 
\sum_{ {\bf{q}} \sigma^{'} }
\left( -A^{\dagger}_{ {\bf{p}} + {\bf{q}}/2 \sigma }
({\bf{q}}\sigma^{'})
 + A_{ {\bf{p}} + {\bf{q}}/2 \sigma^{'} }
(-{\bf{q}}\sigma) \right)
\mbox{      }T_{ {\bf{q}} }({\bf{p}}) } 
\mbox{        }R({\bf{p}} \sigma)
\]
\[
 = -\sum_{ \sigma^{'} }[A^{\dagger}
_{ {\bf{p}} + {\bf{q}}/2 \sigma }({\bf{q}}\sigma^{'}),
A_{ {\bf{k}} \sigma_{1} }({\bf{q}}\sigma_{2})]
T_{ {\bf{q}} }({\bf{p}}) \mbox{        }c_{ {\bf{p}} \sigma }
\]
\begin{equation}
 = \delta_{  {\bf{k}},  {\bf{p}} + {\bf{q}}/2  }
\delta_{ \sigma_{1}, \sigma }
n_{F}({\bf{k}}-{\bf{q}}/2)(1-n_{F}({\bf{k}}+{\bf{q}}/2)) \mbox{    }
\mbox{        }T_{ {\bf{q}} }({\bf{p}}) \mbox{        }c_{ {\bf{p}} \sigma }
\end{equation}

\[
[c_{ {\bf{p}} \sigma },
 A^{\dagger}_{ {\bf{k}} \sigma_{1} }({\bf{q}}\sigma_{2})]
 = e^{ 
\sum_{ {\bf{q}} \sigma^{'} }
\left( -A^{\dagger}_{ {\bf{p}} + {\bf{q}}/2 \sigma }
({\bf{q}}\sigma^{'})
 + A_{ {\bf{p}} + {\bf{q}}/2 \sigma^{'} }
(-{\bf{q}}\sigma) \right)
\mbox{      }T_{ {\bf{q}} }({\bf{p}}) } 
 A^{\dagger}_{ {\bf{k}} \sigma_{1} }({\bf{q}}\sigma_{2})
\mbox{        }R({\bf{p}} \sigma)
\]
\[
 - A^{\dagger}_{ {\bf{k}} \sigma_{1} }({\bf{q}}\sigma_{2})
 e^{ 
\sum_{ {\bf{q}} \sigma^{'} }
\left( -A^{\dagger}_{ {\bf{p}} + {\bf{q}}/2 \sigma }
({\bf{q}}\sigma^{'})
 + A_{ {\bf{p}} + {\bf{q}}/2 \sigma^{'} }
(-{\bf{q}}\sigma) \right)
\mbox{      }T_{ {\bf{q}} }({\bf{p}}) } 
\mbox{        }R({\bf{p}} \sigma)
\]
\[
 = \sum_{ \sigma^{'} }
[A_{ {\bf{p}} - {\bf{q}}/2 \sigma^{'} }({\bf{q}}\sigma),
A^{\dagger}_{ {\bf{k}} \sigma_{1} }({\bf{q}}\sigma_{2})]
T_{ -{\bf{q}} }({\bf{p}}) \mbox{        }c_{ {\bf{p}} \sigma }
\]
\begin{equation}
 = \delta_{  {\bf{k}},  {\bf{p}} - {\bf{q}}/2  }
\delta_{ \sigma, \sigma_{2} }
n_{F}({\bf{k}}-{\bf{q}}/2)(1-n_{F}({\bf{k}}+{\bf{q}}/2)) \mbox{    }
\mbox{        }T_{ -{\bf{q}} }({\bf{p}}) \mbox{        }c_{ {\bf{p}} \sigma }
\end{equation}
 In the sea-boson method, unlike in the usual bosonization in 1d, the momentum
 space description is easier than the real space description. Now we have
 to compute $ R({\bf{p}}\sigma) $ by making contact with the free theory.
 This is proving to be harder than
 expected since the commutator of the exponent with the other exponent
 does not commute with either exponent, hence we may not use the
 truncated version of the Baker-Hausdorff theorem.
 If one insists on computing the full dynamical propagator which is needed
 for obtaining the density of states as a function of the energy,
 then perhaps a simple route may suffice. It goes by the name
 `serendipitous surmise'.

\subsection{ A Serendipitous Surmise }

Many years ago the author had a conversation with then the student
now Prof. A.H. Castro-Neto where the latter suggested that maybe
the field operator is simply given by,
\begin{equation}
\psi({\bf{x}}) \approx e^{-i \sum_{ {\bf{q}} } e^{ i{\bf{q}}.{\bf{x}} }
X_{ {\bf{q}} } } \sqrt{ \rho^{0} }
\label{SEREN}
\end{equation}
 where by definition $ X_{ {\bf{q}} } =
 i {\bf{q}} \cdot {\bf{j}}(-{\bf{q}})/({\bf{q}}^{2}N^{0}) $
 and $ \rho^{0} = N^{0}/V $.  Later he realised that if we choose
 $ [X_{ {\bf{q}} }, X_{ {\bf{q}}^{'} }] = 0 $ as is in fact
 mandatory \footnote{ Since strictly speaking it is the conjugate to $ \rho $
 as defined by the line integral of the ratio of the current and density
 that enters; these commute amongst themselves. }, then fermion commutation
 rules are not obeyed. However one may take solace in the fact that at least
 one commutation rule does come
 out right namely $ [\psi({\bf{x}}), \rho_{ {\bf{q}} }] = 
e^{ i{\bf{q}}.{\bf{x}} } \psi({\bf{x}}) $. A refinement over this ansatz
 was attempted in our earlier work\cite{Setlur1} by introducing
 an additional phase functional of the density linear in the density and
 this was also inadequate since by now the author knows that $ \Phi $ 
 there when computed was imaginary when it was postulated to be real.
 A compromise was also suggested that involved multiplying and dividing
 by the free propagator and using the exact version in the numerator and
 the bosonized free propagator in the denominator. This trick though
 repugnant to most, gives us an anomalous exponent
 of the Luttinger liquid as we shall see below. However, this anomalous
 exponent is off by a factor of two from the exact one obtained by
 Mattis and Lieb.
 From Eq.(~\ref{SEREN}) we may write,
\[
G({\bf{x}}-{\bf{x}}^{'}) = \left< \psi^{\dagger}({\bf{x}}^{'}) 
\psi({\bf{x}}) \right> 
\approx \left< e^{i \sum_{ {\bf{q}} } 
\left( e^{ i{\bf{q}}.{\bf{x}}^{'} }
 - e^{ i{\bf{q}}.{\bf{x}} } \right)
X_{ {\bf{q}} } } \right>
\mbox{        }
\rho^{0}
\]
\begin{equation}
=  e^{- \sum_{ {\bf{q}} } 
\left( 1 - cos[{\bf{q}}.({\bf{x}}-{\bf{x}}^{'})] \right)
\left< X_{ {\bf{q}} } X_{ -{\bf{q}} } \right> }
\mbox{        }
\rho^{0}
\end{equation}
 Again using the trick outlined in our
 earlier work\cite{Setlur1} we may write,
\begin{equation}
G({\bf{x}}-{\bf{x}}^{'}) = G_{0}({\bf{x}}-{\bf{x}}^{'}) \mbox{      }
e^{- \sum_{ {\bf{q}} } 
\left( 1 - cos[{\bf{q}}.({\bf{x}}-{\bf{x}}^{'})] \right)
[\left< X_{ {\bf{q}} } X_{ -{\bf{q}} } \right>
-\left< X_{ {\bf{q}} } X_{ -{\bf{q}} } \right>_{0}] }
\end{equation}
 Here $ G_{0}({\bf{x}}-{\bf{x}}^{'}) $ is the propagator
 obtained from elementary considerations. In one dimension, we may see
 that $ \left< X_{ {\bf{q}} } X_{ -{\bf{q}} } \right> \approx
 k^{2}_{F} S(q) /(q^{2}N^{0}) $. The structure factor
 $ S_{0}(q) = |q|/(2k_{F}) $ for the interacting case. 
 For the interacting case we have, $ S(q) = (v_{F}/v_{eff}) \mbox{ }S_{0}(q) $.
\[
G(x-x^{'}) = G_{0}(x-x^{'}) \mbox{      }
e^{ -\int^{\infty}_{0} dq \mbox{       } 
\frac{ 1 - cos[q.(x-x^{'})] }{|q|} 
\left( \frac{ v_{F} }{2v_{eff}} - \frac{1}{2} \right) }
\]
\begin{equation}
\sim  G_{0}(x-x^{'}) \mbox{      } \left( \frac{1}{|x-x^{'}|} \right)^{\gamma}
\end{equation}
 where $ \gamma = \frac{ v_{F} }{2v_{eff}} - \frac{1}{2} $.   
 This exponent is exactly one half of the exponent 
 obtained by Mattis and Lieb.
 What is even worse is, we have shown in an
 earlier preprint 
  that when applied to the X-ray edge problem, we obtain
 the well-known results of Mahan in one dimension but not in higher dimensions.
 Thus it would appear that there is something amiss in the expression
 for the field operator.
 Nevertheless it may be a quick and easy 
 way of getting the density of states that gives the right qualitative physics
 if not the right exponents.

\end{document}